

\font\titolino=cmbx10
\font\tsnorm=cmr10
\font\tscors=cmti10

\font\tscorsp=cmti9
\magnification=1200

\hsize=148truemm
\hoffset=10truemm
\parskip 3truemm plus 1truemm minus 1truemm
\parindent 8truemm
\newcount\notenumber

\def\PRD{{\tscors Phys. Rev. D }}
\def\NP{{\tscors Nucl. Phys. }}

\def\PLB{{\tscors Phys. Lett. B }}

\def\MPLA{{\tscors Mod. Phys. Lett. A  }}
\def\CQG{{\tscors Class. Quantum Grav. }}
\def\ANP{{\tscors Ann. Physics} (N.Y.) }
\def\mp{M_{pl}^2}
\def\wh{wormhole}
\def\whs{wormholes}
\def\Wh{Wormhole}
\def\Whs{Wormholes}
\def\ks{Kan\-to\-wski\--Sa\-chs}
\def\kl{Kon\-to\-ro\-vich\--Le\-be\-dev}
\def\sc{\scriptstyle}

\def\note{\advance\notenumber by 1 \footnote{$^{\the\notenumber}$}}
\def\ref#1{\medskip\everypar={\hangindent 2\parindent}#1}
\def\beginref{\begingroup
\bigskip
\leftline{\titolino References.}
\nobreak\noindent}
\def\endref{\par\endgroup}
\def\beginsection #1. #2.
{\bigskip
\leftline{\titolino #1. #2.}
\nobreak\noindent}

\def\beginack
{\bigskip
\leftline{\titolino Acknowledgments}
\nobreak\noindent}

\nopagenumbers
\rightline{}
\rightline{\tscors June 1994}
\rightline{Ref. SISSA 92/94/A}
\vskip 20truemm
\centerline{\titolino QUANTUM ELECTROMAGNETIC WORMHOLES AND}
\bigskip
\centerline{\titolino GEOMETRICAL DESCRIPTION OF THE ELECTRIC CHARGE}
\vskip 15truemm
\centerline{\tsnorm Marco Cavagli\`a}
\bigskip
\centerline{\tscorsp SISSA-ISAS, International School for Advanced
Studies, Trieste, Italy}
\centerline{\tscorsp and}
\centerline{\tscorsp INFN, Sezione di Torino, Italy.}
\vfill
\centerline{\tsnorm ABSTRACT}
\begingroup\tsnorm\noindent
I present and discuss a class of solutions of the Wheeler-de Witt
equation describing wormholes generated by coupling of gravity to the
electromagnetic field for Kantowski-Sachs and Bianchi I spacetimes. Since
the electric charge can be viewed as electric lines of force trapped in a
finite region of spacetime, these solutions can be interpreted as the
quantum corresponding of the Ein\-stein\--Ro\-sen\--Mis\-ner\--Whee\-ler
electromagnetic geon.
\vfill
\leftline{PAC(s) NUMBERS: 04.60.-m, 04.60.Ds, 04.60.Kz.}
\medskip
\hrule
\noindent
Mail Address:
\hfill\break
SISSA-ISAS, International School for Advanced Studies
\hfill\break
Via Beirut 2-4, I-34013 Miramare (Trieste)
\hfill\break
Electronic mail: 38028::CAVAGLIA or CAVAGLIA@TSMI19.SISSA.IT
\endgroup
\eject
\footline{\hfill\folio\hfill}
\pageno=1
\beginsection 1. Introduction.
\Whs\ are classical or quantum solutions for the gravitational field
describing a bridge between smooth regions of spacetime. In the classical
case they are instantons describing a tunneling between two distant
regions [1-7]; conversely, in the quantum theory \whs\ are solutions of
the Wheeler-De Witt (WDW) equation [8,9] that reduce to the vacuum wave
function for large three-geometries [10,11].

The existence of microscopic \whs\ may have significant effects at
low-energy scales (see for instance [12]) so much time has been devoted
to look for particular \wh\ solutions. In this paper I derive and discuss
quantum \whs\ generated by the electromagnetic field. Electromagnetic
\whs\ are very appealing for several reasons. For instance, the
anisotropic nature of the electromagnetic field prevents spatially
homogeneus and isotropic solutions of the field equations, so in order to
describe quantum \wh\ solutions we must consider minisuperspace models of
dimension greater than one. As we shall see below, electromagnetic \whs\
have properties that differ strongly from the properties of \whs\
generated through coupling to other fields, as the scalar field. Finally,
the electromagnetic field is a gauge field and it is very important to
study \whs\ generated by gauge fields because the latter constitute a
fundamental ingredient of the modern field theory.

I do not discuss here the most general form of the electromagnetic field
but I limit myself to a particular ansatz (for the general discussion see
[13]). I consider the ansatz used in ref. [6,7] in the contest of General
Relativity and String Theory. Hence, the quantum solutions I find here
correspond to the solutions found in [6,7]. In this case the solutions of
the WDW equation represent the quantum analogue of a classical
electromagnetic geon [14,15]. In ref. [6] it was shown that a Lorentzian
macroscopic observer looking at the \wh\ sees an apparent electric charge
$Q$ even though physical charges are absent. Indeed, the electric field
extends smootly through the mouths of the \wh\ so the observer in the
asymptotic region interprets the electric flux as due to an apparent
charge in the origin. The spacetime is described by a
Reissner-Nordstr\"om type solution with mass $M=0$. An Euclidean \wh\ is
joined at the naked singularity via a change of signature. The solutions
of the WDW equation that I derive in this paper correspond to the
classical solution discussed above; as we will see, the full quantum
treatment avoids the problems related to the change of signature.

The outline of the paper is the following: in the next section I
introduce briefly the WDW equation and the \wh\ wave functions. In the
third section I derive and discuss \wh\ solutions of the WDW equation
both for Bianchi I and \ks\ models. The classical-to-quantum
correspondence for the \ks\ case is also discussed. Finally, we state our
conclusions.
\beginsection 2. WDW Equation and \Wh\ Wave Functions.
Here we discuss briefly the WDW equation and the asymptotic behaviour
that identifies the \wh\ wave functions. Let us consider a Riemannian
four-dimensional space $\Omega$ with metric $g_{\mu\nu}$ and topology
$R\times H$, where $H$ is a three-dimensional compact hypersurface. The
line element can be cast in the form (see for instance [14]):
$$ds^2=g_{\mu\nu}dx^\mu dx^\nu=
(N^2+N_iN^i)d\tau^2+2N_idx^id\tau+h_{ij}dx^idx^j,\eqno(2.1)$$
where $N$ represents the lapse function, $N^i$ is the shift vector and
$h_{ij}$ ($i,j=1,..3$) is the metric tensor of $H(\tau~=~const)$. The
action reads
$$S=\int_\Omega d^4x\sqrt g\biggl[-{\mp\over 16\pi}{}^{(4)}R+
{\cal L}(\phi)\biggr]-{\mp\over 8\pi}
\int_{\partial\Omega}d^3x\sqrt h({\bf K}-{\bf K_0}).\eqno(2.2)$$
Here $M_{pl}$ is the Planck mass, $g=~$det$~g_{\mu\nu}$, ${}^{(4)}R$ is
the curvature scalar and ${\cal L}(\phi)$ represents the Lagrangian
density of a generic matter field $\phi$. The boundary term is required
by unitarity [16]; ${\bf K}$ is the trace of the second fundamental form
${\bf K}_{ij}$ of $H$ and $\bf{K_0}$ is that of the asymptotic
three-surface embedded in flat space. The latter contribution must be
introduced if one requires the space to be asymptotically flat.

Using (2.1), (2.2) becomes
$$S=\int_{\Omega} d^4x\bigl(\pi^{ij}\dot h_{ij}+\pi_{\phi}\dot\phi-NH_0
-N^iH_i\bigr)+\hbox{(surface terms)},\eqno(2.3)$$
where $\pi^{ij}$ and $\pi_\phi$ are respectively the conjugate momenta to
$h_{ij}$ and $\phi$; $H_0$ and $H_i$ are the Hamiltonian generators:
$$\eqalignno{&H_0={16\pi\over\mp}{\cal H}_{ijkl}\pi^{ij}\pi^{kl}+
{\mp\over16\pi}\sqrt h{}^{(3)}R+{\cal H}_0(\phi),&\hbox{(2.4a)}\cr
&H_i=-2\pi^j{}_{i|j}+{\cal H}_i(\phi).&\hbox{(2.4b)}\cr}$$
Here ${\cal H}_0$ and ${\cal H}_i$ are the Hamiltonian generators
of the matter field and
$${\cal H}_{ijkl}={1\over 2\sqrt
h}(h_{ik}h_{jl}+h_{il}h_{jk}-h_{ij}h_{kl})\eqno(2.5)$$
is the metric of the superspace. $N$ and $N^i$ are Lagrange multipliers
that impose the constraints
$$\eqalignno{&H_0=0,&\hbox{(2.6a)}\cr
&H_i=0.&\hbox{(2.6b)}\cr}$$
The quantization can be achieved identifying the classical quantities
$\pi^{ij}$ in (2.4) with the operators
$$\pi^{ij}\rightarrow -\biggl({\mp\over 16\pi}\biggr)
{\delta\over\delta h_{ij}}\eqno(2.7)$$
and analogously for the classical momenta of the matter fields. So using
(2.7), (2.6a) becomes the WDW equation:
$$\Biggl[{\cal H}_{ijkl}{\delta^2\over\delta h_{ij}
\delta h_{kl}}+\sqrt h{}^{(3)}R+{16\pi\over\mp}{\cal H}_0(\phi)
\Biggr]\Psi(h_{ij},\phi)=0.\eqno(2.8)$$
Here ${\cal H}_0(\phi)$ is a quantum operator and $\Psi$ represents the
wave function of the system. In (2.8) there are factor ordering problems
that we disregard for the moment. From (2.6b) we obtain the momentum
constraint equations:
$$\Biggl[-2\biggl({\mp\over 16\pi}\biggr)\biggl[{\delta\over
\delta h_{ij}}\biggr]_{|j}+{\cal H}_i(\phi)\Biggr]\Psi(h_{ij},\phi)=0.
\eqno(2.9)$$
Eqs. (2.9) imply that the wave function is invariant under
diffeomorphisms, i.e., $\Psi$ is only a functional of the three-geometry
and not of the particular three-metric $h_{ij}$.

Now, we may introduce \wh\ wave functions. Following Hawking and Page [11]
we define quantum \whs\ as non singular solutions of the WDW equation and
constraints (2.9) that for large three-metrics reduce to the vacuum wave
function. The latter is defined by a path integral over all
asymptotically Euclidean metrics ${\cal C}$ (not necessarly $R^4$,
depending on topology of the space) and matter fields vanishing at
infinity:
$$\Psi(\tilde h_{ij},\tilde\phi)=\int_{\cal C}{\cal D}g_{\mu\nu}{\cal D}\phi
e^{-S[g,\phi]}.\eqno(2.10)$$
Following Garay (see [10]), we may calculate the path integral (2.10).
Here $\cal C$ represents the class of three-metrics and matter fields
which satisfy the conditions
\medskip
\line{\hbox to 40truemm{\hfill}
\hbox to 40truemm{$h_{ij}(x,\tau)=\tilde h_{ij}$,\hfill}
\hbox to 40truemm{$\phi(x,\tau)=\tilde\phi$,\hfill}\hfill (2.11a)}
\bigskip
\line{\hbox to 40truemm{\hfill}
\hbox to 40truemm{$\phi(x,\infty)=0$,\hfill}
\hbox to 40truemm{$h_{ij}(x,\infty)=h_\infty$.\hfill}\hfill (2.11b)}
\medskip\noindent
$h_\infty$ is the asymptotic metric which identifies the space
with minimal gravitational excitation. Evaluating (2.10) in the
asymptotic limit, we obtain the asymptotic behaviour of the \wh\ wave
function [10]
$$\Psi(\tilde h_{ij})\approx\exp\biggl[\int
d^3x\pi^{ij}h_{ij}\biggr]_{\tau}.\eqno(2.12)$$
Finally, (2.11) and (2.12) are the boundary conditions which identify the
\wh\ wave functions.
\beginsection 3. Bianchi I and Kantowski-Sachs models.
Let us consider $H=T^3(\chi,\theta,\varphi)$, where $T^3$ represents the
three-torus. The line element (2.1) can be written [17,18]:
$$ds^2=\rho^2\bigl[N^2(\tau)d\tau^2+a^2(\tau)d\chi^2+b^2(\tau)d\theta^2
+c^2(\tau)d\varphi^2\bigr],\eqno(3.1)$$
where $\rho^2=2/M^2_{pl}\pi^2$ and $\chi$, $\theta$ and $\varphi$ are
defined in the interval $[0,2\pi[$.

The electromagnetic Lagrangian in (2.2) is ${\sc{1}\over{\sc
2}}F\wedge{}^*F$. We choose the electric field along the $\chi$ direction
[6,7]:
$${\bf A}={1\over\sqrt{2\pi^3}}A(\tau)d\chi.\eqno(3.2)$$
Clearly, analogous results can be obtained choosing $\theta$ or $\varphi$
directions. Of course, (3.2) is not the most general ansatz for the
electromagnetic field (for a most general treatment, see [13]).
Substituting (3.1) and (3.2) in (2.3) and integrating over the spatial
variables, the action becomes (we neglect surface terms)
$$S=\int_\tau^\infty Nd\tau\Biggl[{1\over N^2}f_{\alpha\beta}
\dot q^\alpha\dot q^\beta\Biggr],\eqno(3.3)$$
where $q=(A,a,b,c)$ and dots represent differentiation with respect
to $\tau$.
$$f_{\alpha\beta}=\pmatrix{&2bc/a&0&0&0\cr
&0&0&-c~&-b~\cr
&0&-c~&0&-a~\cr
&0&-b~&-a~&0\cr}\eqno(3.4)$$
is the four-dimensional metric of the minisuperspace whose line element
reads
$$d{\cal S}^2=2\biggl[{bc\over
a}dA^2-cdadb-bdadc-adbdc\biggr].\eqno(3.5)$$
The asymptotic behaviour of the \wh\ wave functions (2.12) is
$$\Psi_{\rm as}(a,b,c)\approx\exp\Biggl[-{2\over N}{d\over
d\tau}(abc)\biggr|_{\tau}\Biggr].\eqno(3.6)$$
Note that the asymptotic behaviour of the \wh\ wave functions does not
depend on the structure constants of the three-dimensional isometry group
$G$ generating a homogeneous three-surface $H$ [17,18], so (3.6) holds
for the all Bianchi and \ks\ models. Here we will calculate (3.6) for
Bianchi I space. The calculation for the \ks\ space is analogous and I
will give only the result.

To calculate (3.6) we need the classical Euclidean equations of motion.
Varying (3.3) with respect to the scale factor and electromagnetic
potential, we obtain in the gauge $N=1$:
$$\eqalignno{&{\ddot a\over a}+{\dot a\dot b\over ab}+{\ddot b\over
b}=-{\dot A^2\over a^2},&\hbox{(3.7a)}\cr
&{\ddot c\over c}+{\dot c\dot a\over ca}+{\ddot a\over
a}=-{\dot A^2\over a^2},&\hbox{(3.7b)}\cr
&{\ddot b\over b}+{\dot b\dot c\over bc}+{\ddot c\over
c}={\dot A^2\over a^2},&\hbox{(3.7c)}\cr
&4\dot A{bc\over a}=K,&\hbox{(3.7d)}\cr}$$
where $K$ is the conjugate momentum to $A$. The Hamiltonian constraint is
$${\dot a\dot b\over ab}+{\dot c\dot a\over ca}+
{\dot b\dot c\over bc}={\dot A^2\over a^2}.\eqno(3.8)$$
Let us assume that (3.1) becomes asymptotically an Euclidean Kasner
universe of indices $p$, $q$ and $r$ when $\tau\rightarrow\infty$ (see
for instance [14]). Evaluating the curvature tensor in this limit we find
that (3.1) is flat when $p,q,r<1$ or when $p=1$, $q=r=0$ (and cyclic
permutations); in the latter case the topology is $R^2\times T^2$. For
these values there are no gravitational excitations in the asymptotic
region. When $p=q=r=1$ or $r<p=q=1$ or $q,r<p=1$, $q,r\not=0$ (and cyclic
permutations) the Riemann tensor is non vanishing and then the minimal
asymptotic gravitational excitation is different from zero.

The equation (3.7d) fixes the asymptotic behaviour of the EM
potential to be
$$A(\tau)\approx\left\{\matrix{&\tau^{p-q-r+1}&p-q-r\not=1,\cr\cr
&\log(\tau)&p-q-r=1.\cr}\right.\eqno(3.9)$$
(2.11b) implies $q+r>p+1$ so the previous conditions on $p$, $q$ and
$r$ reduce to
\medskip
\line{\qquad\hbox to 20 truemm{1)\hfill}\hbox to 40 truemm{$p,q,r<1$,\hfill}
\hbox to 40 truemm{$p+1<q+r<2$;\hfill}\hfill}
\medskip
\line{\qquad\hbox to 20 truemm{2a)\hfill}\hbox to 40 truemm{$q=1$,\hfill}
\hbox to 40 truemm{$p<r<1$;\hfill}\hfill}
\medskip
\line{\qquad\hbox to 20 truemm{2b)\hfill}\hbox to 40 truemm{$r=1$,\hfill}
\hbox to 40 truemm{$p<q<1$;\hfill}\hfill}
\medskip
\line{\qquad\hbox to 20 truemm{3)\hfill}\hbox to 40 truemm{$q=1$,\hfill}
\hbox to 20 truemm{$r=1$,\hfill}
\hbox to 20 truemm{$p<1$.\hfill}\hfill}
\medskip
These relations rule out \whs\ with asymptotic topology $R^2\times T^2$.
{}From the equations of motion for the scale factors (3.7) and from the
Hamiltonian constraint (3.8) we easily deduce that the stress
energy-momentum tensor $T_{\mu\nu}$ of the electromagnetic field is
asymptotically
$$T_{\mu\nu}\approx\tau^{-2(q+r)}.\eqno(3.10)$$
The stress energy-momentum tensor must vanish for $\tau\rightarrow\infty$
as $G_{\mu\nu}$ or faster, so we obtain a further condition on $q$ and
$r$:
$$q+r\ge 1.\eqno(3.11)$$
Since the electromagnetic potential along the $\chi$ direction does not
break the isotropy along $\theta$ and $\varphi$ directions, we can put
$b=c$ in the classical equations (3.7,8) and 1-3 reduce to ${{\sc
1}\over{\sc 2}}\le q=r\le 1$, $p<1$. When $q=r={{\sc 1}\over{\sc 2}}$ we
must consider non-vacuum equations. Putting $a=(\alpha\tau)^p$ and
$b=c=(\alpha\tau)^q$ and substituting in (3.7) and (3.8) we obtain
\medskip
\line{\hbox to 30 truemm{\hfill}
\hbox to 50truemm{$a=(\alpha\tau)^{-1/2}$,\hfill}
\hbox to 50truemm{$b=c=(\alpha\tau)^{1/2}$,\hfill}\hfill (3.12a)}
\medskip\noindent
where $\alpha\equiv\omega/2=iK/2$, and
\medskip\noindent
\line{\hbox to 30 truemm{\hfill}
\hbox to 50truemm{$a=(\alpha\tau)^{-1/3}$,\hfill}
\hbox to 50truemm{$b=c=(\alpha\tau)^{2/3}$.\hfill}\hfill (3.12b)}

In both cases the space is asymptotically flat so the minimal
gravitational excitation is asymptotically zero. Choosing
$\alpha=\omega/4$ in (3.12b) and substituting $a$, $b$ and $c$ in (3.6),
we find
$$\eqalign{\Psi_{\rm as}\approx &e^{-\omega
a(a^\rho b^\sigma c^\lambda)/2},\cr
\rho=\sigma+\lambda\hbox to 12truemm{\hfill}&
\hbox to 12truemm{\hfill or \hfill}\hbox to 12truemm{\hfill}
\rho=2(\sigma+\lambda)-1,\cr}\eqno(3.13)$$
where $\sigma$ and $\lambda$ are two arbitrary positive parameters. The
\wh\ wave functions can behave for large three-geometries essentially in two
different ways according to the asymptotic three-metric. This occurs
because the asymptotic region with minimal gravitational excitation is
not unique and does not have the topology $R^3\times S^1$ as it happens,
for instance, to the \ks\ model. Hence, the \wh\ wave functions which for
large three-geometries behave as
$$\Psi_{\rm as}\approx e^{-\omega a^2\sqrt{bc}/2}\eqno(3.14)$$
represent Riemannian spaces asymptotically of the form (3.12a) or (3.12b)
and correspond to \whs\ joining two asymptotically flat regions with
metric (3.12a) and (3.12b).

Now we are able to find \wh\ solutions for Bianchi I model. Choosing the
Hawking-Page prescription for the factor ordering [11], the WDW equation
(2.8) can be cast in the form
$$\Delta\Psi(a,b,c,A)=0,\eqno(3.15)$$
where $\Delta$ is the Laplace-Beltrami covariant operator in the
minisuperspace:
$$\eqalign{\Delta={2\over c}\partial_a\partial_b+{2\over b}\partial_a\partial_c
+{2\over a}\partial_b\partial_c-{a\over bc}\partial_a^2&-{b\over ac}
\partial_b^2-{c\over ab}\partial_c^2+\cr
&+{1\over bc}\partial_a-{1\over ac}
\partial_b-{1\over ab}\partial_c-{a\over bc}\partial_A^2.\cr}\eqno(3.16)$$
A set of solutions of (3.15) is
$$\Psi(a,b,c,A;p,k,\omega)={1\over\sqrt{bc}}a^{i(p+k)/2}b^{ip/2}c^{ik/2}
K_{i\sqrt{pk}}(\omega a)e^{i\omega A}\eqno(3.17)$$
where $p$, $k$ and $\omega$ are real constants and $K$ is the modified
Bessel function. The wave functions (3.17) oscillate an infinite number
of times approaching the origin. They are damped for large values of $a$
but not for large $b$ and $c$ because they oscillate in the $b-c$ plane.
This feature is not surprising, since we have chosen a electromagnetic
field that ``lives'' in the one-sphere with radius $a$, so its dynamics
does not depend on $b$ and $c$ [19]. Using the variables $\log b$ and
$\log c$, we see that the oscillating factors $b^{ip/2}$ and $c^{ik/2}$
look formally as $e^{i\omega A}$. So, as far as the dynamics of the \wh\
is concerned, the scale factors $b$ and $c$ behave essentially as matter
fields and toghether with $A$ determine the extent of the \wh\ mouth
[19]. Note that for real values of $\omega$ there is a real flux of the
electric field through any closed $T^2(\theta,\phi)$ surface due to the
electric field along $\chi$. The physical meaning of this flux will be
more clear in a moment. Finally, the factor $1/\sqrt{bc}$ is eliminated
by the measure in the integral when we deal with matrix elements, so the
probability density $\sqrt f|\Psi|^2$ is finite everywhere.

Considering a linear combination of wave functions (3.17), we can find
regular \wh\ wave functions. Let us put $p=\xi^2k$, where $\xi$ is a real
number, and take the Fourier transform [20]. We
obtain (see [21], for instance):
$$\eqalign{\Psi(a,b,c,A;\mu,\xi,\omega)&=\int dk e^{i\mu
k}\Psi(a,b,c,A;k,\xi,\omega)\cr
&={1\over\sqrt{bc}}e^{i\omega A}e^{-\omega
a\cosh{[\log(a^{\sigma+\lambda}b^\sigma c^\lambda)+\mu]}},}\eqno(3.18)$$
where $\sigma=\xi/2$ and $\lambda=1/2\xi$. Wave functions (3.18)
represent Riemannian spaces asymptotically of the form (3.12a) and are
the analogue for the electromagnetic field of the solutions found by
Campbell and Garay in ref. [22] for a scalar field in \ks\ space.
Solutions (3.18) are again eigenfunctions of the operator
$\partial/\partial A$, so there is a real flux through any closed
$T^2(\theta,\phi)$ surface as for (3.17). The physical interpretation of
(3.18) must take some care since the regularity at small three-geometries
means that the space closes regularly there, so the electromagnetic flux
cannot go through the \wh\ throat. We will see below the physical meaning
of this feature.

Choosing $\xi=1$ we obtain
$$\Psi(a,b,c,\mu,\omega)={1\over\sqrt{bc}}e^{i\omega A}e^{-\omega
a\cosh{[\log(a\sqrt{bc})+\mu]}}.\eqno(3.19)$$
Now, let us put for simplicity $\mu=0$. (3.19) can be cast in the form
$$\Psi(a,b,c,A;\omega)={1\over\sqrt{bc}}e^{-\omega a^2\sqrt{bc}/2}
e^{-\omega/2\sqrt{bc}}e^{i\omega A}\eqno(3.20)$$
which coincides with a \kl\ transform (see [21]) of (3.17) with respect
to the index $k=p$. Using a different type of \kl\ transform we can find
a further solution
$$\Psi(a,b,c,A;\omega)={1\over\sqrt{bc}}
\biggl(a^2\sqrt{bc}-{1\over\sqrt{bc}}\biggr)e^{-\omega a^2\sqrt{bc}/2}
e^{-\omega/2\sqrt{bc}}e^{i\omega A}.\eqno(3.21)$$
The asymptotic behaviours of (3.20) and (3.21) suggest us the use in the
WDW equation of the new variables
$$\eta^2=\omega a^2\sqrt{bc}/2,\hbox to
30truemm{}\xi^2=\omega/2\sqrt{bc}.\eqno(3.22)$$
Recalling (3.15) we obtain
$$\Psi_n(\eta,\xi,A;\omega)=\psi_n(\eta+\xi)
\psi_n(\eta-\xi)\xi^2e^{i\omega A},\eqno(3.23)$$
where $\psi_n(x)$ is the harmonic wave function of order $n$:
$$\psi_n(x)={1\over\bigl(2^n
n!\sqrt\pi\bigr)^{1/2}}H_n(x)e^{-x^2/2}.\eqno(3.24)$$
Solutions (3.20) and (3.21) correspond (apart from normalization factors)
to $\Psi_0$ and $\Psi_1$. As in the previous set of solutions, we have a
non zero flux even if the wave functions are regular for small
three-geometries. This important property makes (3.19) and (3.24) very
different from other quantum \whs\ known in literature. As we have seen,
{\tscors both for singular and regular wave functions} there is a real
flux. For solutions (3.17) this is not surprising, because one can
imagine the flux coming out or going into the singularity at the origin.
However, for solutions (3.23) where can the flux go?

To answer to this question and shed light on the physical interpretation
of the solutions, we have to consider the structure of the (Euclidean)
electromagnetic field [23]. As we have said before, the ansatz (3.2)
represents a purely electric field along the $\chi$ direction, i.e. a
electromagnetic field whose dynamics is confined in the one-sphere with
radius $a$. Conversely, the asymptotic behaviour of the wave function for
large three-geometries and the throat depend on $b$ and $c$. Hence, the
dynamics of the electromagnetic field is decoupled from the dynamics of
the \wh\ and the flux of the electric field must coincide for regular and
non-regular wave functions [19].

Even though there are no physical charges in the field equations, the
observer in the asymptotically flat region measures a real finite flux
and sees an {\tscors apparent} charge in the origin. Thus the geometry
must be non trivial. Since the solutions (3.24) describe asymptotically
flat spaces, the electromagnetic field is confined in a finite region. We
can conclude that (3.24) describe the quantum analogue of a
electromagnetic geon because the charge can be seen as an electric field
trapped in a finite region of space, without any source.

This interpretation holds also for the \ks\ model. Moreover, in the latter
case we can compare the quantum results to the classical ones, since
the classical \wh\ solution corresponding to the ansatz (3.2) is known.
Let us now discuss briefly the \ks\ model. The quantum \wh\ solutions for the
ansatz (3.2) have been found in [24]. They are
$$\Psi(a,b,A;\nu,\omega)=K_{i\nu}(\omega a)K_{i\nu}(4ab)
e^{\pm i\omega A}.\eqno(3.25)$$
The wave functions (3.25) are singular in the origin, where they oscillate an
infinite number of times. For $4ab<|\nu|$ the wave functions
oscillate (Lorentzian region), conversely for $4ab>|\nu|$ they are
exponentially damped (Euclidean region). The asymptotic behaviour
corresponds to the flat $R^3\times S^1$ space. Hence, (3.25) represent
quantum \whs\ with throat $4ab=|\nu|$ joining two asymptotically flat
regions with topology $R^3\times S^1$. Note that the size of the throat
does not depend on the scale factor $a$ nor on the parameter $\omega$.

Multiplying (3.25) by a factor $\nu\tanh(\pi\nu)$ and taking
the \kl\ transform, (see [21]) we obtain:
$$\Psi(a,b,A;\omega)={2\sqrt{b\omega}\over\omega+4b}
e^{-a(\omega+4b)}e^{\pm i\omega A}.\eqno(3.26)$$
We can easily verify that (3.26) is regular everywhere and its asymptotic
behaviour for $b\rightarrow\infty$ coincides with the behaviour of
(3.25). As in Bianchi I model, we have again regular solutions but a non
zero electromagnetic flux, so (3.26) must describe a non trivial
topology. Indeed, since there are no physical sources for the
electromagnetic field, the Gauss law should imply a vanishing flux
through closed surfaces around the origin. We conclude that (3.26)
represent a quantum electromagnetic geon. The latter interpretation is
supported by the classical solution corresponding to this model. As shown
in [6] for the classical Einstein-Maxwell theory, a macroscopic observer
measures an apparent electric charge $Q$ even though physical charges are
absent. Since the spacetime is not trivial, the electric field extends
smootly beyond an Euclidean \wh\ throat joined to the Lorentzian
isometric regions via a classical change of signature. The Lorentzian
geometry is described by the metric:
$$ds^2
=-\biggl(1+{Q^2\over R^2}\biggr)dT^2+\biggl(1+{Q^2\over R^2}\biggr)^{-1}
dR^2+R^2d\Omega_2^2\eqno(3.27)$$
where the radial coordinate $R$ ranges in $]~0,\infty~[$. The line element
(3.27) is of a Reissner-Nordstr\"om type and the joining occurs in the
naked singularity at $R=0$. Indeed, the latter can be ``expanded'' and
continued analitically in the Euclidean space, where the solution is
everywhere regular. This Euclidean region describes the \wh. The wave
functions (3.26) are the quantum analogue of this picture. In the full
quantum treatment we avoid the problems of the classical case, namely the
``ad hoc'' change of signature: solutions (3.26) are regular everywhere
and thus describe a geometry that closes regularly. In the classical
theory, the lines of force of the electric field are convergent at $R=0$,
where the Lorentzian spacetime becomes singular. So we need a little
``trick'', namely the transition from a Lorentzian to an Euclidean region.
In the quantum theory, we have solutions regular everywhere and no
``tricks'' are necessary. We stress that these results depend strongly on
the ansatz (3.2) chosen for the electromagnetic field. Naturally,
identical conclusions can be drawn for the Bianchi I model.
\beginsection 4. Conclusions.
In this paper we have found and discuss a class of solutions of the WDW
equation. These solutions represent \whs\ generated by coupling gravity
and the electromagnetic field. They have topology $R\times T^3$ or
$R\times S^1 \times S^2$. We have considered a particular ansatz for the
electromagnetic field, namely a purely electric field along a $S^1$
section. These solutions can be interpreted as the quantum corresponding
of a electromagnetic geon. The electric charge is viewed as electric
lines of force trapped in a finite region of spacetime. This
interpretation has been discussed in detail for the $R\times S^1 \times
S^2$ topology and the correspondence with the classical Einstein-Maxwell
theory has been explored.
\beginack
I am very grateful to Luis J. Garay for very fruitful suggestions about the
interpretation of the solutions and useful remarks. I also thank Vittorio de
Alfaro for interesting discussions and suggestions and Fernando de Felice
for addressing us to the geon topic.
\vfill\eject
\beginref

\ref [1] S.B. Giddings and A. Strominger, \NP B {\bf 306} (1988) 890.

\ref [2] J.J. Halliwell and R. Laflamme, \CQG {\bf 6} (1989) 1839.

\ref [3] D.H. Coule, and K. Maeda, \CQG {\bf 7} (1990) 955.

\ref [4] S.W. Hawking, \PLB {\bf 195} (1987) 337.

\ref [5] A. Hosoya and W. Ogura, \PLB {\bf 225} (1989) 117.

\ref [6] M. Cavagli\`a, V. de Alfaro and F. de Felice, \PRD {\bf 49}
(1994) 6493.

\ref [7] M. Cadoni and M. Cavagli\`a, {\tscors Cosmological and Wormhole
Solutions in low-energy effective String Theory}, Report No: SISSA
75/94/A, INFNCA-TH-94-11, hep-th/9406053.

\ref [8] B. DeWitt, \PRD {\bf 160}, (1967) 1113.

\ref [9] J.A. Wheeler, in: {\tscors Battelle Rencontres: 1967 Lectures in
Mathematics and Physics}, eds. C. DeWitt and J.A. Wheeler, W.A. Benjamin, New
York, 1968.

\ref [10] L.J. Garay, \PRD {\bf 44} (1991) 1059.

\ref [11] S.W. Hawking and D.N. Page, \PRD {\bf 42} (1990) 2655.

\ref [12] S. Coleman, \NP B {\bf 310} (1988) 643.

\ref [13] H.F. Dowker, \NP B {\bf 331} (1990) 194.

\ref [14] Misner C.W., K.S. Thorne and J.A. Wheeler, {\tscors
Gravitation}, 1973, W. H. Freeman and Company, New York.

\ref [15] C.W. Misner and J.A. Wheeler, \ANP {\bf 2} (1957) 525.

\ref [16] S.W. Hawking, in {\tscors General Relativity, an Einstein
Centenary Survey}, eds. S.W. Hawking and W. Israel, Cambridge University
Press, Cambridge, 1979.

\ref [17] M.P. Ryan and L.C. Shepley, {\tscors Homogeneus Relativistic
Cosmologies}, Princeton University Press, Princeton, New Jersey, 1975.

\ref [18] R.T. Jantzen, in {\tscors Cosmology of the Early Universe},
eds. by L.Z. Fang and R. Ruffini, World Scientific, Singapore, 1984.

\ref [19] We are very grateful to L.J. Garay for suggesting us this
interpretation.

\ref [20] For the physical meaning of the Fourier transform in
minisuperpace models, see for instance L.J. Garay, \PRD {\bf 48} (1993)
1710 and G. A. Mena Marugan, {\tscors Bases of Wormholes in Quantum
Cosmology}, Report No: CGPG-94/4-4, gr-qc/9404042; {\tscors Wormholes as
Basis for the Hilbert Space in Lorentzian Gravity} Report No:
IMAFF-RC-04-94, CGPG-94/5-2, gr-qc/9405027.

\ref [21] Bateman Manuscript Project, {\tscors Tables of Integral
Transforms}, 1954, Mc. Graw--Hill Book Company, New York; Vol. I and II.

\ref [22] L.M. Campbell and L.J. Garay, \PLB {\bf 254} (1991) 49.

\ref [23] For the classical Einstein-Maxwell theory in the Euclidean
space, see for instance: D. Brill, Report No: UMD 93-038, gr-qc/9209009,
to appear in {\tscors Proceedings of Louis Witten Festschrift}, World
Scientific, Singapore.

\ref [24] M. Cavagli\`a, {\tscors Quantum Wormholes in the
Kan\-to\-wski\--Sa\-chs Spacetime}, Report No: SISSA 74/94/A, to appear
in \MPLA.

\endref
\vfill
\bye